\documentstyle[11pt,aas2pp4,flushrt]{article}
\singlespace
\newcommand{\etal}{{et al.~}}
\newcommand{\eg}{{e.g.,} \thinspace}
\newcommand{\ie}{{i.e.,} \thinspace}
\newcommand{\Msol}{\mbox{M$_\odot$}} 
\def\uv{\hbox{$U\!-\!V$}}
\def\vi{\hbox{$V\!-\!I$}}
\def\vk{\hbox{$V\!-\!K$}}
\def\br{\hbox{$B\!-\!R$}}

\slugcomment{submitted to ApJ, January 28, 1998}
\begin{document}

\today

\title{Photometric Evolution of Galaxies in Cosmological Scenarios}
\lefthead{Contardo \etal}
\righthead{Photometric evolution of galaxies}

\author{Gertrud Contardo\altaffilmark{1,2}, Matthias Steinmetz\altaffilmark{1,3}, and 
Uta Fritze--von Alvensleben\altaffilmark{4}}
\altaffiltext{1}{Max-Planck-Institut f\"ur Astrophysik, Postfach 1523, 85740 Garching, Germany}
\altaffiltext{2}{Present address: European Southern Observatory, Karl-Schwarzschild-Str.~2, 
85748 Garching, Germany}
\altaffiltext{3}{Steward Observatory, University of Arizona, Tucson, AZ 85721, USA}
\altaffiltext{4}{Universit\"atssternwarte G\"ottingen, Geismarlandstra{\ss}e 11, 37083
G\"ottingen, Germany}
\centerline{gcontard@eso.org,msteinmetz@as.arizona.edu,ufritze@uni-sw.gwdg.de}

\begin{abstract}

The photometric evolution of galaxies in a hierarchically clustering
universe is investigated.  The study is based on high resolution numerical
simulations which include the effects of gas dynamics, shock heating,
radiative cooling and a heuristic star formation scheme. The outcome of
the simulations is convolved with photometric models which enables us to predict
the appearance of galaxies in the broad band colors $U$, $B$, $V$, $R$, $I$ and $K$.  We
demonstrate the effect of the mutual interplay of the hierarchical build-up of
galaxies, photometric evolution, k-correction, and intervening absorption on the
appearance of forming disk galaxies at redshift one to three. We also discuss to
what extend the numerical resolution of current computer simulations is
sufficient to make quantitative predictions on surface density profiles and
color gradients.

\end{abstract}


\section{Introduction}
Over the past few years, the advent of a new generation of large
telescopes and the superb imaging quality of the refurbished
Hubble-Space-Telescope have changed the way we can look at the formation and
evolution of galaxies. Galaxies are routinely identified at redshift
$z\approx 1$ and also objects at redshifts $z\approx 3$ have been optically
resolved (Steidel \etal 1996). Theoretical work has been guided by the
hierarchical clustering hypothesis. In this scenario, the {\sl Cold Dark Matter} (CDM) 
model is probably the best known example, structures form as halos of
progressively larger mass merge, virialize and form new higher mass
objects. Thus, the galaxy formation process can be modeled from ab-initio
conditions and model predictions can be rigorously compared to observational
data. Two complementary techniques have been shown to be especially
powerful. 

In so-called semi-analytical studies (Kauff\-mann, White \& Guiderdoni 
1993; Cole \etal 1994), phenomenological recipes are used to model the
accumulation of gas at the center of dark matter halos and its transformation
into stars.  The involved
free parameters are calibrated by matching present day properties of galaxies,
\eg the properties of the Milky Way or the Tully-Fisher relation.  These
techniques are comparably cheap as far as computation is concerned and thus allow a large coverage
of the parameter space. The major strength of the semi-analytical approach lies
in modeling statistical properties of the whole galaxy population as, \eg 
the luminosity function of galaxies. Its disadvantage is that little can be
said on details of the galaxy formation process or the exact structure and
kinematics of galaxies. 

Numerical hydrodynamical simulations which follow the
formation of individual galaxies in detail
fill in that gap. Indeed, advanced numerical simulations include all the relevant
physical processes: gravitation, gas dynamics, adiabatic heating and cooling,
hydrodynamic shocks, radiative cooling and star formation (Katz 1992; Navarro \&
White 1993; Steinmetz \& M\"uller 1994, 1995; Evrard, Summers \& Davis
1994, Tissera, Lambas \& Abadi, 1997). However, the required numerical resolution ($\approx 1\,$kpc) together
with the demanding numerical algorithms and the limited computational power of
the present generation of supercomputers constrain current simulations to small box
sizes, \ie each simulation only creates very few galaxies. Thus hydrodynamical
simulations can deal with the details of the galaxy formation process, but can
make only limited predictions on statistical properties.

Numerical simulations have also helped identifying some of the major problems of the
hierarchical clustering hypothesis: Since cooling times scale inversely with density, the
dissipative collapse of gas should have been more efficient at high
redshift because the dark matter halos present at that time were denser, as was the universe as a whole. Thus gas cools and settles into
rotationally supported disks already when the first level of the hierarchy collapses.
Consequently, models predict an overabundance of low mass galaxies as well as
disk galaxies which for a given circular velocity are too massive (White \& Rees 1978).
A related problem is the size of present day disk galaxies: Due to the frequent
merging angular momentum is efficiently transported from the baryons to the
dark-matter halo resulting in disk galaxies which are too concentrated compared
to present day disk galaxies (Navarro, Frenk \& White 1995; Navarro \&
Steinmetz 1997). Feedback processes due to star formation, like 
stellar winds or supernovae, have repeatedly been envisioned to solve this over-cooling problem
(\eg Dekel \& Silk 1986). However, a simulation which includes the effect of
star formation {\sl and}\/ which is consistent with observations has not been
performed yet.

In this paper we merge the approach of numerical simulations with traditional methods of
spectrophotometric synthesis. This hybrid scheme enables us to perform  an even more
realistic comparison between model predictions and actual observations.
The structure of the paper is the following: In
section 2 details of the numerical simulations and the star formation scheme are
provided and the spectrophotometric modeling is presented.  
In section 3 we apply these methods to model the
appearance  of high redshift galaxies as they form in CDM like universes. In
section 4 we describe to what extend current simulations may even allow us to
analyze the detailed structure of galaxies at high and low redshift.  
Section 5 summarizes the results and draws our
conclusions.

\section{Methods}
\subsection{Numerical simulations}

The simulations were performed using GrapeSPH (Steinmetz 1996), a hybrid
scheme of a smoothed particle hydrodynamics (SPH) approach with a direct
summation N-body routine which uses the hardware N-body integrator GRAPE. The code is
specially designed to follow a three component system of gas, stars, and dark
matter and has high resolution in space and time due to the use of individual
smoothing lengths and time steps. It includes the effects of pressure gradients,
adiabatic heating and cooling, shocks and radiative cooling. 

In the simulations presented here, a star formation scheme similar to the one in
Steinmetz \& M\"uller (1994, 1995, henceforth SM1 and SM2;
see also Katz 1992) is used: Star formation is modeled by
creating new collisionless ``star particles'' in regions where the gas is
locally Jeans unstable and where the cooling timescale is shorter than the local
dynamical time scale.  The orbits of these newly formed star particles are subsequently
followed in a self-consistent fashion, assuming that they are only affected by
gravitational forces. Young star particles devolve energy and metal enriched
mass to their surrounding gas, an effect that mimics the energy and mass input
by supernovae and evolving stars into the ISM. The supernova energy is added to
the thermal energy of the surrounding gas. Input in kinetic energy (Navarro \&
White 1993) is not directly implemented but is described by an efficiency
parameter which is of the order of a few per cent, an assumption that is
justified since we preferably concentrate on the formation of luminous galaxies
like the Milky Way.

\subsection{Initial conditions}
We use  idealized initial conditions as proposed by Katz (1991)
and discussed in more detail in SM2: We consider an isolated sphere on which small scale
fluctuations according to a CDM power spectrum are superimposed. 
To incorporate the effects of fluctuations with longer wavelengths, the density of the
sphere has been enhanced and rigid rotation corresponding to a spin parameter
of $\lambda=0.08$ has been added. The simulation is identical to that presented
in SM1 and SM2  where a disk galaxy similar to the Milky
Way forms. These idealized initial conditions are fairly realistic for the
early formation epochs ($z>2$) when the sphere itself has not yet collapsed.
The late formation epoch, however, is dominated by the collapse of the sphere
and hence proceeds more smoothly than under more realistic
conditions. Due to the smooth infall at low redshifts the simulation
does not suffer from the angular momentum problem. Furthermore, this model
allows for very high numerical  resolution (spatial resolution 1\,kpc, mass
resolution $10^7$\,\Msol) at comparably low computational cost. 

\subsection{Simulation outcome}

\begin{figure}[ht]
\epsscale{1}
\plotone{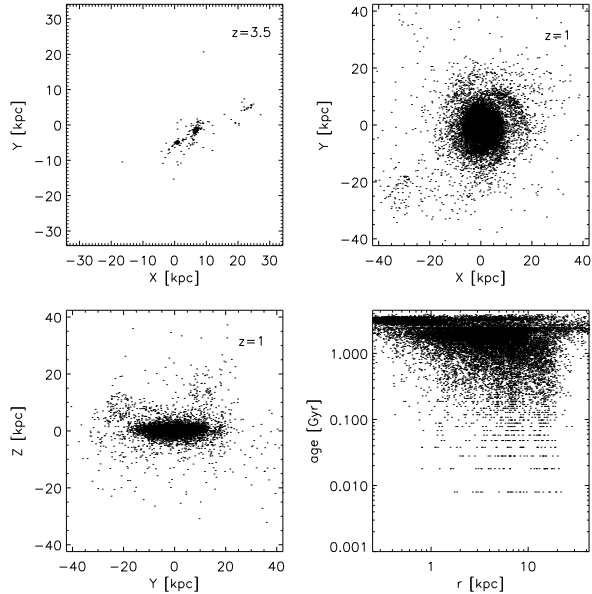}
\caption{\label{particles}Upper left: distribution of star particles at redshift 
$z=3.5$. The sidelength of the plots corresponds to 100 pixels of the HST/WFPC2. 
Upper right: face-on view of the galaxy at redshift $z=1$. Lower left:
edge-on view of the galaxy at redshift $z=1$. Lower right: age of star particles
at $z=1$ as a function of their distance to the galaxy center.}
\end{figure}

At any redshift, the simulations provide the spatial distribution of star particles. Each 
star particle corresponds to a population of about one million stars which have
formed in a burst, \ie the whole galaxy  is a superposition of several ten 
thousands of mini starbursts. Each of the star particles is described by its age (time passed
since creation) and its metallicity. 

The actual distribution of star particles is shown in figure \ref{particles}.
Details on the formation history of the model galaxy, of its kinematical
properties as well as of its chemical evolution can be found in SM1 and
SM2. Here we concentrate on some features important for the discussion later in
this paper.  At redshift 3.5, two protogalactic clumps can be seen which are
going to merge within the following $\approx 10^8$ years. At lower redshifts,
the stars originating in these two clumps can be found in the bulge of the
galaxy. At $z=1$, the model galaxy already resembles a Milky Way type galaxy
with a bulge-to-disk ratio of 1:3 in mass (SM2).  In the lower right panel of
figure \ref{particles}, we plot the age of the star particles as a function of
the distance to the galaxy center. The central 1-2\,kpc are dominated by old
stars with an age between 3 and 4\,Gyr. The majority of these star particles
has a metallicity of above solar (SM1). There is little offset between the age of
the oldest stars (3.9\,Gyr) and the age of the universe at that redshift
(4.6\,Gyr). This feature is common to any hierarchically clustering universe in
which the first stars form in small clumps which collapse at high redshifts. 
The actual value here is
only an upper limit since (i) the limited numerical resolution cannot follow
star formation in clumps with less then $\approx 10^6\,$\Msol and (ii) the
amount of substructure is underrepresented due to the idealized initial
conditions.  At larger radii ($r>2\,$kpc), younger stars with an age of about
1\,Gyr and below are dominating which reflects ongoing star formation in the
gaseous disk.  Due to the vacuum boundary conditions, the evolution of the
galaxy is fairly quiescent at redshifts below one. Gas in the disk is
successively transformed into stars. Due to the still insufficient numerical
resolution, however, encounters between disk stars and dark matter particles act
as a kinematic heating term. While at $z=1$ the velocity dispersion of disk
stars is still small ($\approx 10\,$km/sec), the stellar disk is successively
heated up between $z=1$ and $z=0$ up to velocity dispersions near 100\,km/sec
at $z=0$.  We therefore focus our analysis mainly on redshift one, when
the amount of artificial heating is still negligible.

\subsection{Spectral synthesis}
As mentioned above, each star particle, which has a typical mass of a few
million solar masses, corresponds to a population of several million stars which
have been formed at the same time $t_f$ from a gas cloud with a metallicity of
$Z$.  A star particle can thus be considered as a population of stars which have
formed in a burst-like manner.  The luminosity evolution of such a burst is then
followed by an evolutionary spectral synthesis model (Fritze-von Alvensleben \& Gerhard
1995;  Fritze-von Alvensleben \& Burkert 1995; Einsel \etal 1995): 
The evolution of this instantaneously
formed population of stars is followed in the Hertzsprung--Russell diagram (HRD)
using results of stellar evolution theory.  Sets of stellar tracks (Maeder
1990) for five
different metallicities ($Z = 10^{-4}$, $10^{-3}$, $4\times10^{-4}$, $10^{-2}$,
$4\times10^{-2}$) and for 27 stellar masses in the range of $0.15\,\Msol$ to
$60\,\Msol$ have been compiled by Einsel \etal (1995).  A Miller-Scalo (1979)
initial mass function (IMF) is assumed with upper and lower mass limits of
$60\,\Msol$ and $0.15\,\Msol$, respectively.  The spectrum of an individual star 
burst is synthesized from Kurucz's (1992) library of stellar spectra, 
luminosities in various filter bands are obtained by convolution with filter
response functions (see also Leitherer \etal 1996, Fritze-von Alvensleben \& Kurth 1996). 

\begin{figure}[ht]
\plotone{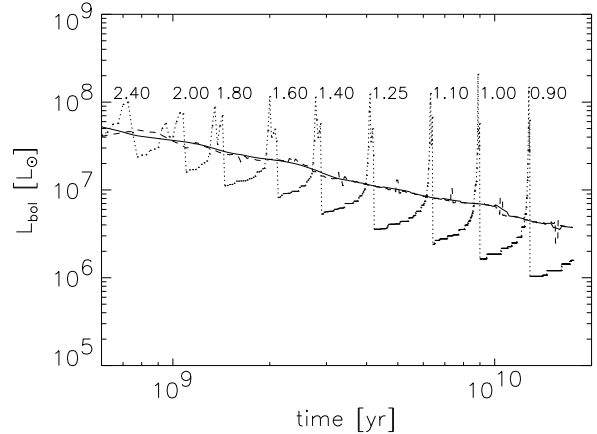}
\caption[]{\label{glatt-unglatt} Luminosity evolution of a stellar population
formed within $2.5\times10^7~\mathrm{years}$ with metallicity $Z =
1\times10^{-4}$.  The dotted line shows the evolution as computed with the
original code. The luminosity indicated by the dashed line is calculated with
the averaged stellar evolution tracks and smoothed subsequently (solid
line). The numbers right of the peaks assign the mass $m_i$ in \Msol of the star that
reaches the red giant branch at that time.}
\end{figure}

The discrete set of stellar masses for which stellar evolution tracks are
available together with the burst-like
star formation history can cause some problems for the late time evolution of the
population.  In the original implementation, all stars within a mass interval
$\left[m_{i-1/2};m_{i+1/2}\right]$ have been assigned to the temperature and
luminosity of the stellar track of a star of mass $m_i$. For masses below
$2\,\Msol$, the lifetime on the red giant branch (a few $10^8\,$yr) is, however,
short compared to the difference in the lifetime of stars on neighboring
evolutionary tracks ($10^9$\,yr). This causes all stars within a given mass
interval to
synchronously enter the red giant branch. Since the luminosity increases
hundredfold, and since every mass interval contains a substantial fraction of
the total mass of the population (a few per cent), pronounced spikes in the
luminosity can be seen (figure \ref{glatt-unglatt}, dotted line). Such
a discreteness dominated behavior is unknown for extended star formation
histories (as e.g. in models of spiral galaxies), since these spikes are
convolved (and thus smoothed) with the star formation history.

\begin{figure}[ht]
\plotone{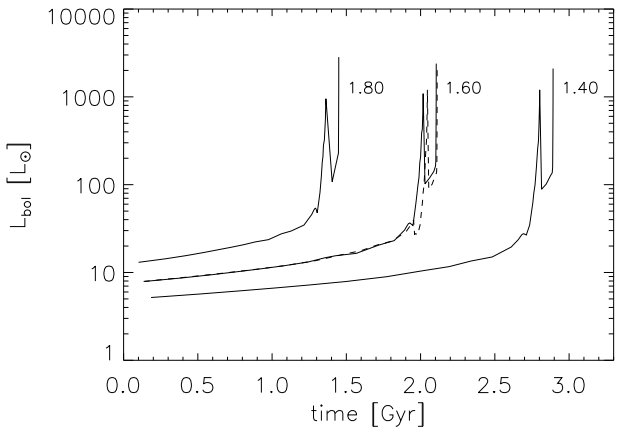}
\caption{\label{interpol_track}Comparison of the luminosity evolution of three
stars of mass $M= 1.4\,\Msol$, $1.6\,\Msol$ and $1.8\,\Msol$, respectively. The
dashed line has been obtained by interpolating between the $M= 1.4\,\Msol$ and
$1.8\,\Msol$ track as described in the text.}
\end{figure}

These discreteness effects in the luminosity evolution of burst-like formed
populations are circumvented by methods which are based on 
isochrones (see, \eg Bruzual \& Charlot 1993). We propose an alternative method that
avoids the construction of isochrones. The luminosity evolution of
an individual star is split into two parts, main sequence and post main
sequence. As can be seen in figure \ref{interpol_track}, the post main sequence
evolution between neighboring masses is almost identical, concerning the time
dependence as well as the luminosity evolution. The main sequence 
evolution $l_{\rm MS}\left(t,m\right)$ has also a very similar shape, but
the evolution is faster for the higher masses and the star is also
brighter. These two effects can, however, be compensated by the following ansatz
for the luminosity evolution of a star with mass $m+\Delta m$:
\begin{equation}
l_{\rm MS}\left(t,m+\Delta m\right) = l_{\rm MS}\left(t',m\right)\,
\left(\frac{m+\Delta m}{m}\right)^\alpha \, .
\end{equation}
In this ansatz, $\alpha$ is the exponent of the mass-luminosity relation,
typical values being $\alpha\approx 3$ for masses less than a few \Msol. The
rescaled time $t'$ can be inferred by rescaling $t$ with the ratio of the life
times on the main sequence $\tau_{\rm MS}$, \ie
\begin{equation}\label{aehnlich}
t' = t\, \frac{\tau_{\rm MS}\left(m+\Delta m\right)}{\tau_{\rm MS}\left(m\right)} = 
t\,\left(\frac{m+\Delta m}{m}\right)^\beta
\end{equation}
with $\beta\approx -2$; stars with a higher mass reach a given evolutionary
stage faster. For each mass $m$ within the mass interval
$\left[m_{i-1/2};m_{i+1/2}\right]$ the luminosity evolution can thus be
calculated. The local value of the slope of the mass-luminosity and
mass-lifetime relation, $\alpha$ and $\beta$, can be drawn from neighboring
masses $m_i$. In figure \ref{interpol_track} we demonstrate the success of our
interpolation scheme by comparing the evolution of a star of mass $M= 1.6\,\Msol$
with the evolution as inferred from a star with $M= 1.4\,\Msol$ and $M= 1.8\,\Msol$
using the interpolation scheme above. Differences in the main sequence evolution
are almost undetectable and also the post main sequence evolution differs only
by a very small time shift (less than 10 per cent of the lifetime on the
post-main-sequence).

By averaging these approximated stellar evolution tracks in each mass interval and
by adding up the luminosities according to the initial mass function (IMF) the 
total luminosity of the stellar population is obtained as a function of time.
The luminosity of populations which formed in a single star
burst now evolves much smoother (see figure \ref{glatt-unglatt}, dashed
line). The remaining small wiggles result from discontinuities in the evolution
near the interface of two adjacent mass intervals. We
remove these wiggles by averaging over time intervals of
$1.5\times10^8~\mathrm{years}$, as indicated by the solid line in figure
\ref{glatt-unglatt}.

\begin{figure}[ht]
\plotone{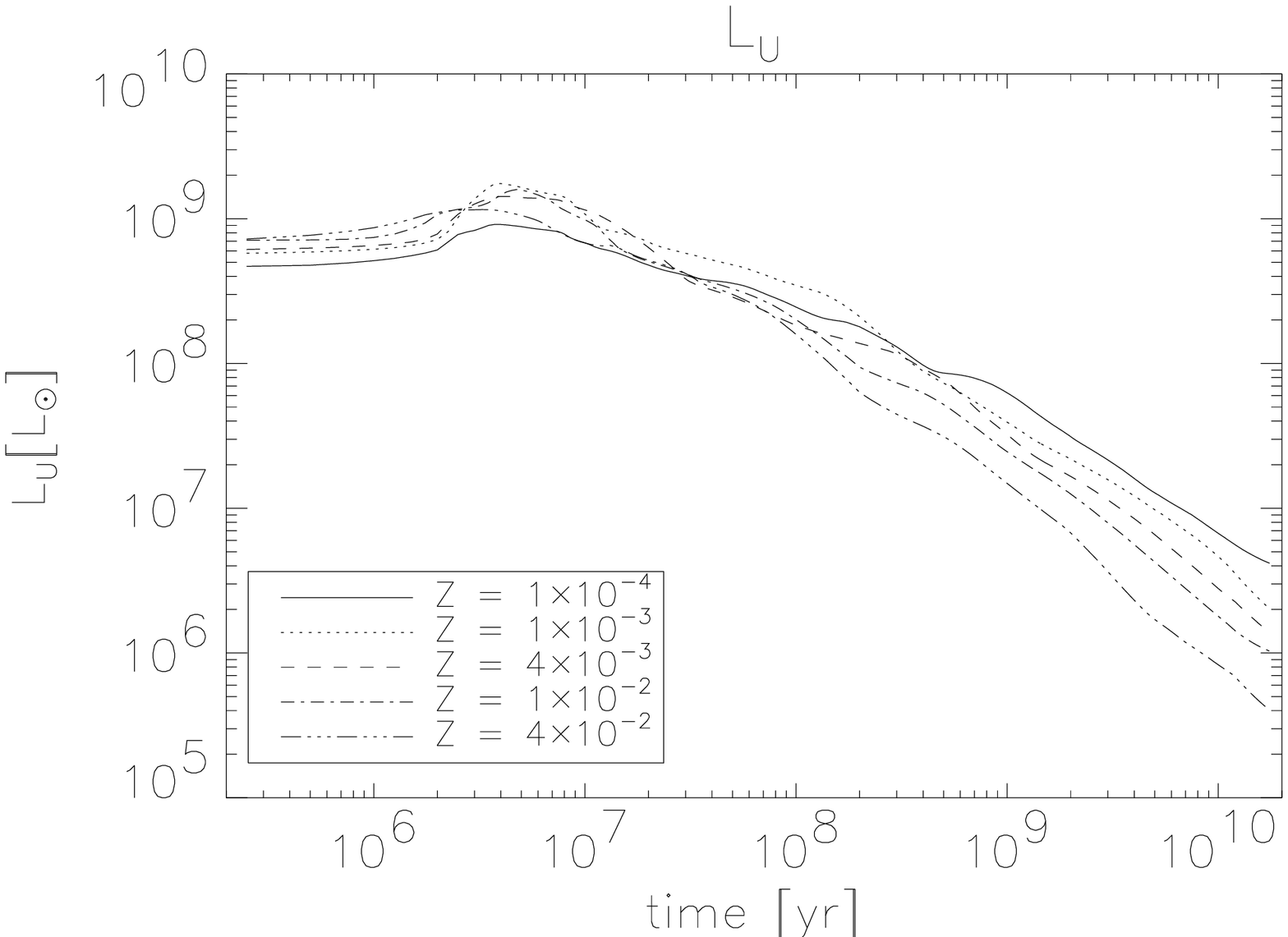}

\plotone{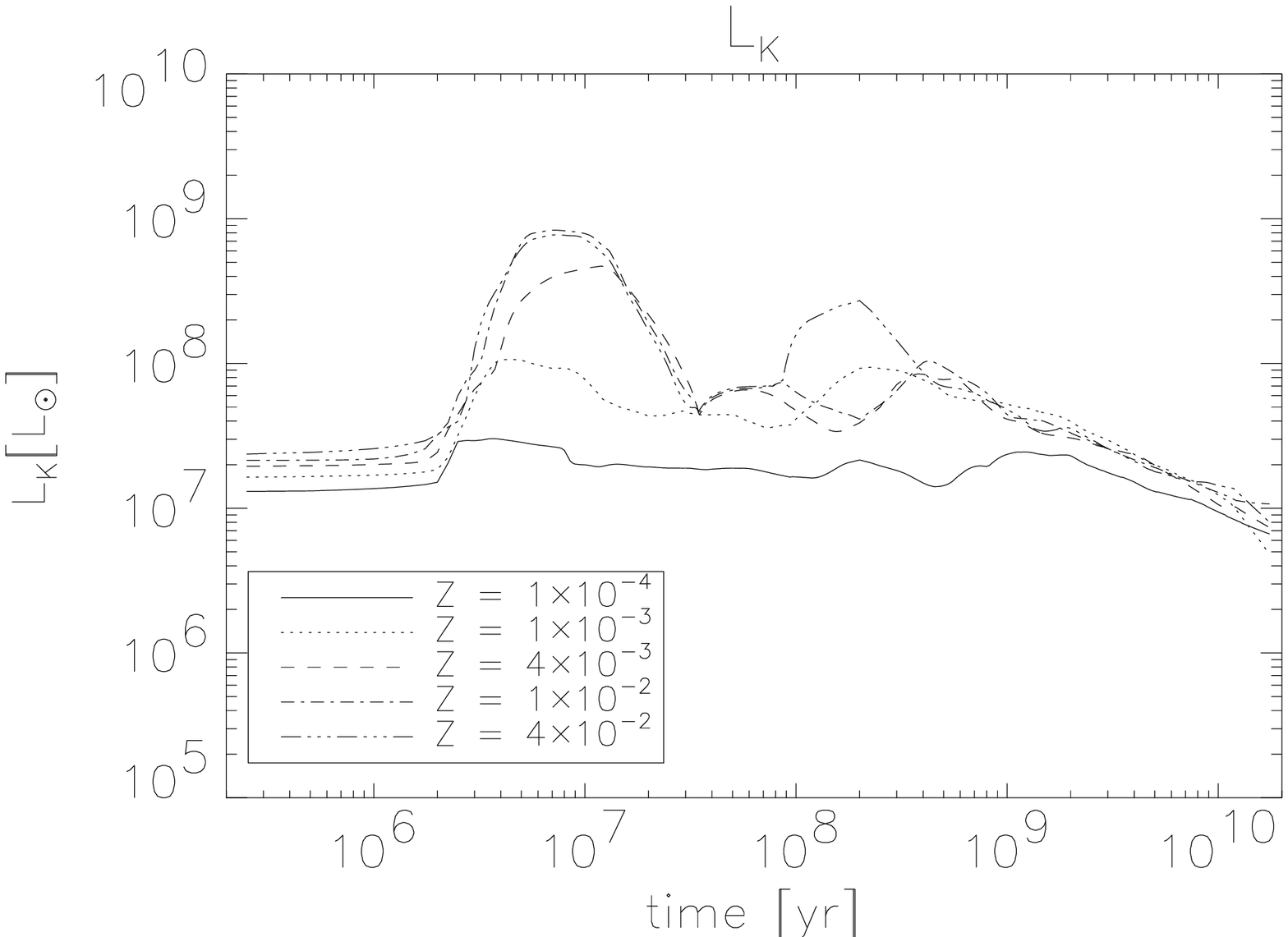}

\caption[]{\label{burstlu}Luminosity evolution in the $U$ and $K$ band of
a single age stellar population for the metallicities $Z = 1\times10^{-4},
1\times10^{-3}, 4\times10^{-3}, 1\times10^{-2}$, and $ 4\times10^{-2}$.  The
total mass of the population is $1 \times 10^7$\,\Msol.}
\end{figure}

The luminosity evolution in the $U$ and $K$ band of a single star burst is shown 
for 5 different metallicities in figure \ref{burstlu}. Data for arbitrary metallicities
$Z$ can then be obtained by interpolation in $\log Z$.

\subsection{k correction}

For apparent magnitudes and colors of galaxies at high redshifts the
modifications due to the redshift of the spectrum have to be taken into
account. 

The cosmological redshift causes the spectral energy distribution
$F_0\left(\lambda\right)$ of a galaxy in its restframe at redshift $z_r$ to be detected
in the observers restframe as $F\left(\lambda\right)$ with:
\begin{eqnarray}
F\left(\lambda,z_r\right) &=& \frac{F_0\left({\lambda}/{\left(1+z_r\right)}\right)}{\left(1+z_r\right)}.
\end{eqnarray}

The apparent magnitude in some ``$X$'' filter band is thus
\begin{equation}\label{appmag}
m_{X} = BDM\left(z_r\right) -
2.5\log\left(\frac{L_{X0}}{\mathrm{L_{X{\sun}}}}\right) +
\mathrm{M_{X{\sun}}} + k\left(z_r\right).
\end{equation}
with the luminosity $L_{X0}$ in the galaxy's rest frame, the bolometric
distance modulus
\[BDM\left(z_r\right) = 5 \log \left({D_L\left(z_r\right)} / {\mathrm{Mpc}}\right) + 25
,\] 
and the luminosity distance (Mattig 1958) 
\begin{eqnarray}
D_L\left(z_r\right) &=&
\frac{c}{H_0 q_0^2}\left\{z_rq_0 + \left(q_0-1\right)\nonumber\right.\\
&& \times \left. \left[-1+\sqrt{\left(2q_0z_r+1\right)}\right]\right\}.
\end{eqnarray}
The {\em k correction} is calculated from the
spectral energy distribution $F\left(\lambda\right)$ and the filter response
function $S_X\left(\lambda\right)$ (Lamla 1982) according to the following
equation (Oke \& Sandage 1968) 
\begin{eqnarray}
k &= &-2.5 \log \left(L_{X{\rm Obs.}}/L_{X0}\right) 
\\
&=& 2.5 \log \frac{\left(1+z_r\right)\int_0^\infty F_0\left(\lambda\right)
S_X\left(\lambda\right)\,d\lambda} {\int_0^\infty
F_0\left(\lambda/\left(1+z_r\right)\right) S_X\left(\lambda\right)\,d\lambda}.
\end{eqnarray}

\subsection{Creating artificial images}

From the luminosity of the set of star particles, artificial images can be
easily generated. The luminosity in a given pass band filter can be converted
into number of photons emitted per unit time per solid angle. This number is
corrected by intervening absorption due to neutral hydrogen using the models of
Madau (1995). By specifying exposure time and pixel size, the number of photons
received by each pixel is calculated. The pixel map then is convolved
with the numerical resolution of the simulation and the sky
background is added. Finally it is convolved with the point-spread function of the
instrument. The resulting array contains the expectation value for each pixel
to receive a photon. This number is then random-sampled. Together with the
detector characteristics (quantum efficiency, readout noise, dark current, gain)
an artificial image can be created which subsequently can be analyzed in the same
way as raw observational data.

\begin{figure}[ht]
\epsscale{0.9}
\plotone{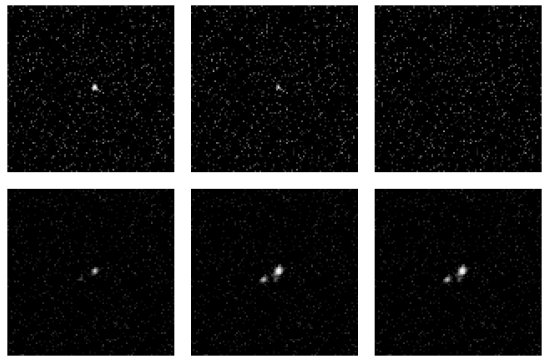}
\caption{\label{ccd1}Artificial HST images of a forming galaxy at $z=3.5$. 
Images are 50 pixels across, ``exposure time'' is 30000 sec. Top row:
$U$-band; bottom row: $I$-band. Left column: rest frame; middle
column: including k-correction; right column:
including k-correction and absorption due to intervening hydrogen.}
\end{figure}

\begin{figure}[ht]
\epsscale{0.9}
\plotone{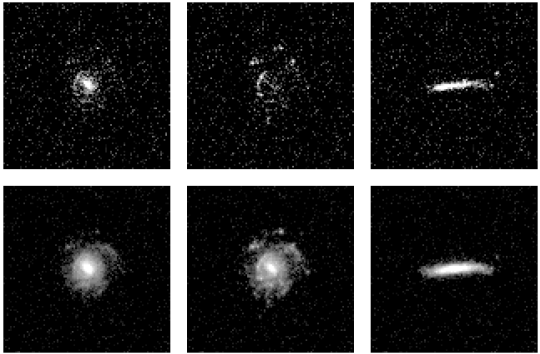}
\caption{\label{ccd2}Artificial HST images of a forming galaxy at $z=1$. 
Images are 50 pixels across, ``exposure time'' is 30000 sec. Top row:
$U$-band; bottom row: $I$-band.  Left column: face on, rest frame; middle
column: face on, including k-correction; right column: edge-on, 
including k-correction}
\end{figure}

\section{Applications}

\subsection{Appearance of high redshift galaxies}

In this paragraph we discuss the appearance of the simulated galaxies in an
exposure as it is typical for HST/WFPC2  observations. Detector characteristics
and sky background have been taken from the WFPC2 handbook and a generous exposure time
of 30000\,s has been specified. For the adopted cosmological model ($\Omega=1$,
$H_0 = 50$\,km\,s$^{-1}$\,Mpc$^{-1}$) and redshift
range ($z=1-4$), the formal numerical resolution of the simulation surpasses the
pixel resolution of the WFPC2 camera by about 50\%.

In figure \ref{ccd1} and \ref{ccd2} we show a $U$ and $I$-band image of the simulated galaxy at
redshifts $z=1$ and $z=3.5$. Each image has a side length of 100
pixels, corresponding to 70\,kpc ($z=3.5$) and 82\,kpc ($z=1$), respectively. 
Grayscales have been chosen logarithmically to cover the interval
between a signal-to-noise of 1 to 30 ($U$ band) and 100 ($I$ band).

\begin{figure}[ht]
\epsscale{0.85}
\plotone{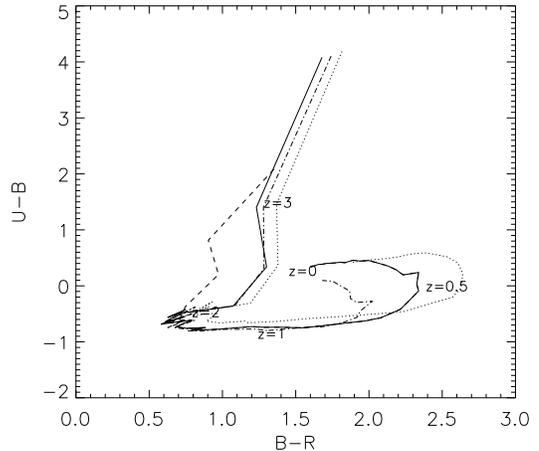}
\caption{\label{stdplot}Evolution of the model galaxy in the \ub\/ vs.~\vr\/  plane. Solid line:
model galaxy, including k-correction and intervening absorption; dashed line:
model galaxy, k-corrected, no absorption; dotted line: E galaxy; dashed-dotted: Sd galaxy
(from M\"oller \etal 1997).}
\end{figure}

In the restframe $U$-band image of the z=3.5 galaxy, only the most massive clump
is visible. It has an apparent magnitude of $m_U=24.8$. Applying the k-correction 
makes the clump barely visible ($m_U=25.7$). At these redshifts, the $U$-band is
centered at wavelengths shortward of $912$\,\AA\/ in the galaxy's restframe where stellar emission is depleted
due to absorption in the stellar atmosphere. Absorption due to intervening
neutral hydrogen suppresses the $U$-band emission of the proto galaxy by further
3.5 mag ($m_U=29.2$) and makes it invisible even in long exposures. 

In the $I$-band, the k-correction has the opposite effect: in its restframe, the
apparent magnitude is $m_I=24.6$ and also the less luminous progenitor is barely
visible. The luminosity of the galaxy is dominated by stars which have formed
within the last few $10^8$\,yr and also massive stars $M>8\,\Msol$ are abundant.
Thus, these lumps strongly emit in the near UV ($\approx 4000$\,\AA). The cosmic 
redshift translates these frequencies to the $I$-band resulting in a negative
k-correction of 1.7 magnitudes ($m_I=22.9$). The galaxy thus has a brightness
compatible to the brightest galaxies in the Steidel \etal~(1996) sample.
Since at $z=3.5$ the $I$-band corresponds to wavelengths longward of $1217\,$\AA,
the luminosity is not affected by intervening absorption.

The restframe $U$-band image of the $z=1$ galaxy consists of a luminous bulge of
stars with a typical age of $3-4$\,Gyrs (see figure \ref{particles}, lower right
panel) and some light from star forming regions
in the disk. At  redshift 1, however, the $U$-band corresponds to restframe
wavelengths near $1800\,$\AA\ where the emission of old stars is small. Thus the
bulge component virtually disappears after applying the k-correction.  Similar
to the $z=3.5$ $I$-band, k-correction enhances the star forming regions in the
disk. Young massive stars emit strongly at restframe wavelengths near 2000\,\AA\
which are redshifted into the $U$-band. Similar effects can be seen in 
the $I$-band image: Since the $I$-band corresponds to a restframe wavelength of 4000\,\AA, also
the light of the old bulge component is clearly visible, although slightly dimmed.
For the disk population the k-correction favors shorter wavelengths at which the
contribution of newly formed stars is larger. Consequently, the k-corrected 
$I$-band image looks more structured and the regions of active star formation are 
more pronounced.

In figure \ref{stdplot} the evolution of the model galaxy in the \ub\ 
vs.~\br\ plane is shown and, for comparison, the evolution of a Sa and Sd galaxy
using traditional spectrophotometric models (see, \eg Bruzual \& Charlot, 1993,
Fritze-von Alvensleben, 1993) is plotted.

\subsection{Surface brightness, color gradients}

\begin{figure}[ht]
\epsscale{0.8}
\plotone{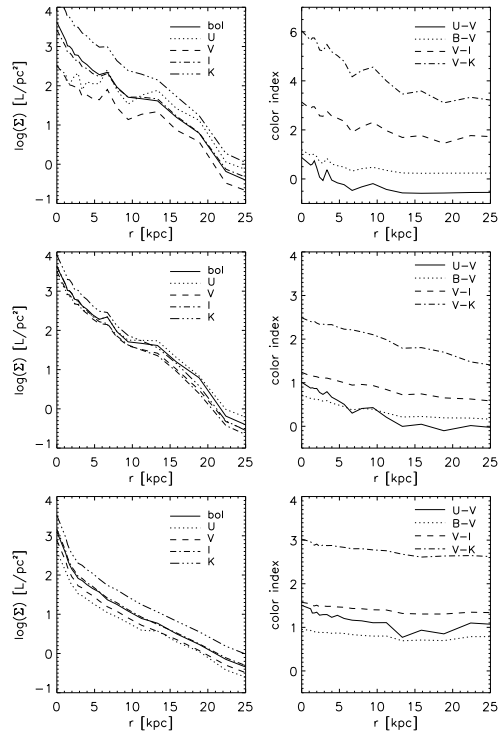}
\caption{\label{color}Left column: Surface density versus radius. The solid line 
corresponds to the bolometric luminosity, the
dashed, dotted, dashed-dotted and dashed-triple-dotted line correspond to the broadband colors
$U$, $V$, $I$ and $K$, respectively. Right column: color indices \uv (solid), \bv(dotted),
\vi (dashed) and \vk (dashed-dotted) versus radius. 
From top to bottom: $z=1$ including k-correction, $z=1$ restframe and $z=0$.}
\end{figure}

In figure \ref{color} we show surface brightness profiles and color
gradients. The radial profile clearly shows an exponential behavior for $r>5\,$kpc
and a steepening (bulge) at smaller radii. The non-exponential bulge component is readily
visible in the $z=0$ plot since the surface brightness of the disk has dimmed by 
about one magnitude. A comparison of the k-corrected (figure \ref{color} top)
galaxy at $z=1$ with its restframe (figure \ref{color} middle) exhibits a
substantial (slight) enhancement of the $K$ ($I$) luminosity (\ie the k-correction
is negative), while the $V$ and $U$ luminosity is dimmed.

At $z=1$ the galaxy exhibits very strong reddening within the inner 5 kpc, about
2 magnitudes in \vk\ and 1 magnitude in the other colors. Except \uv, the drop is 
much weaker in the restframe. This reddening is consistent with the dimming of
the bulge component of the $z=1$ galaxy due to k-correction as described in
section 3.1. The bulge is dominated by old stars (see figure \ref{particles},
lower right panel) emitting only little near 1800\,\AA\ which corresponds to the U
band at $z\approx 1$. The k-correction enhances the $K$ and $I$ luminosity while
decreasing the U and V luminosities and thus results in the strong reddening as can 
be seen in figure \ref{color}. Only a slight color gradient is visible at radii
$r>10\,$kpc in the restframe as well as in the k-corrected profile. Also the
effect of the k-correction is much more moderate, between 0 mag in \bv,
 0.6 mag in \uv\ and 1.5\,mag in \vk. At these radii, stars are still being formed
resulting in a much flatter spectral energy distribution and thus a much weaker
k-correction.

At $z=0$, only very weak color gradients are visible. The bulge differs from the
disk by about 0.1 mag, only in \uv\ a slightly stronger color gradient is
visible. The change of color is certainly much weaker than the observed one of about
0.2 mag in \bv\ and 0.5 mag in \uv\ 
as in galaxies like, \eg Andromeda (Segalovitz 1975, Josey \& Arimoto 1982). 
One reason for this discrepancy is
the neglect of dust absorption. As we will show in the next
paragraph, however, the lack of significant color gradients may also be an artifact
of our star formation scheme and of the vacuum boundary conditions.
 
\begin{figure}[ht]
\epsscale{0.95}
\plotone{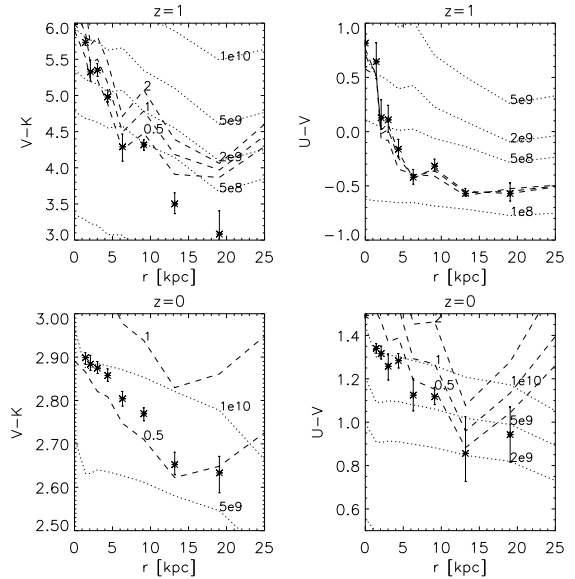}
\caption{\label{colerr}k-corrected \vk\ (left) and \uv\ (right) as a function of radius for the 
model galaxy at redshift $z=1$ (top) and  $z=0$ (bottom). Stars 
correspond to the model galaxy, error bars are $[10;90]$ percentile values based
on a bootstrap analysis (for details, see text). The dashed lines correspond to
a model with constant metallicity  while stellar ages are
taken from the simulation. The dotted lines correspond to a model with constant
stellar ages  while metallicities are taken from the simulation. 
A comparison of the actual values
with the dashed and dotted lines thus demonstrates the effect of metallicity and
age on the color gradient.}
\end{figure}

In figure \ref{colerr} we investigate to what extend the color gradients at $z=1$
and $z=0$ reflect age and metallicity gradients of the stellar disk. First we 
estimate the statistical error of the color indices. For that
purpose we perform a bootstrap analysis, \ie we draw 100 random samples from the
distribution of stars and determine color as a function of radius. We
use the $[10;90]$ percentile values as the lower and upper error limits,
\ie only 10\% of our bootstrapped samples have a color bluer than our lower
error limit and  only 10\% of our bootstrapped
samples have a color redwards of our upper limit. As can be seen, colors
can be measured with a statistical error of about $\pm 0.02$ to $0.05$ mag at
$z=0$. At $z=1$, errors are at the level of about $\pm 0.1$ to $\pm 0.2$.
These errors strongly depend on whether or not stars are actively formed. While
colors are fairly accurate for passively evolving populations, errors are much
larger for regions including young stars. This can be easily seen in the $z=0$ \uv\ 
color at a radius of 12\,kpc. This radial bin contains
very few very young stars. The color thus strongly
depends on how many of these young star particles happen to be included in the
bootstrapped sample.

We also compare the radial color profiles with those obtained assuming a
constant metallicity for the star particles. For the $z=1$ model the color 
near the center is consistent with a metallicity of $Z\approx Z_\odot$, 
while the color at 15\,kpc requires a metallicity much below $0.5\,Z_\odot$. No
constant metallicity model fits the entire color gradient. The metallicity
gradient inferred from the color is consistent with the directly measured
metallicity gradient of $0.05\,$kpc$^{-1}$. The color gradient is not
consistent with the assumption of a constant age. \uv\ near the center requires a 
stellar age of 3\,Gyr, while at radii of 15\,kpc an average age of a few
$10^8$ years seems to provide the best fit, being consistent with the age
distribution shown in figure \ref{particles}. A comparison of  \uv\ and \vk\
also demonstrates that the age distribution at a particular radius  is inconsistent
with the assumption of a constant age: For example, at radii below $r=5$\,kpc
the age which fits \vk\ is more than 2 times larger than the best fitting age
for \uv.

At $z=0$, this picture changes. The color gradient is consistent with a constant 
metallicity of about $0.5\,Z_\odot$ and only a very slight age gradient is
visible.  $Z=0.5\, - 1.0\,Z_\odot$ is the average metallicity of Milky Way 
bulge stars (Mc William \& Rich 199), $Z=0.5\,Z_\odot$ compares well with the
stellar metallicities in Sb models of M\"oller \etal (1997). 
However, the actual metallicity gradient in this model is  much
weaker ($<0.01\,$kpc$^{-1}$). A comparison of the $z=1$ and $z=0$ model thus
provides a further argument for the lack of a significant color gradient at $z=0$
and the too small metallicity gradient. As shown in SM1, the model galaxy also
exhibits a too large fraction of low metallicity stars (the so-called G-dwarf
problem). The reason for the failure to reproduce a
significant metallicity gradient can be found either in the chosen star
formation law or in the vacuum boundary conditions or both. Between $z=1$ and
$z=0$ the disk is dynamically inactive. Only gas is continuously transformed
into stars. Thus each fluid element can be considered as a closed box model. 
The metallicity of a closed box model increases monotonically until the yield is
reached. The yield is, however, identical for each fluid element since the same IMF
has been used. The almost constant metallicity at $z=0$ thus reflects the fact that the
metallicity of each fluid element  has already reached its asymptotic
value. This artifact can be avoided by: (i) decreasing the star formation
efficiency; in that case a metallicity gradient forms since the
metallicity of low density  gas at larger radii does not yet reach its 
asymptotic value; (ii) inflow of gas at redshift $z<1$, \ie non-vacuum boundary 
conditions. We argue that (ii) will be a necessary entity, not only since such a 
late infall is a generic prediction of any hierarchically clustering scenario,
but also because it may help to cure the G-dwarf problem. 

\section{Discussion and conclusions}
We have simulated the formation of a Milky-Way type galaxy in a hierarchically
clustering scenario using idealized initial and boundary conditions. The
simulation outcome has been analyzed with spectrophotometric methods. The most
important results are:

\begin{enumerate}
\item A hybrid scheme of high resolution
numerical
simulations of galaxy formation and spectrophotometric modeling is presented. 
We demonstrated the capabilities
of such an approach. In particular, we
were able to model galaxies with an appearance qualitatively 
similar to observed galaxies at redshifts between $z=0$ and $z=4$. Using the
appearance of $z=1$ and $z=3.5$ galaxies we also could nicely demonstrate the
mutual interplay between the galaxy formation process, k-correction and
intervening absorption.

\item The numerical resolution of present day computer simulations enables us
to study the detailed structure of galaxies as a function of redshift, such 
as surface brightness profiles and color gradients. Numerical resolution is high enough and
statistical errors are sufficienly small to infer whether a significant
metallicity or age gradient of the stellar population are present.

\item The properties of a model galaxy are among others dependent on the
particular star formation scheme, IMF, supernovae feedback. We
are, however, optimistic, that a rigorous comparison of high-resolution
numerically simulated galaxies with the vast amount of detailed observations at
different redshifts may allow us to constrain and tighten the parameter space. Such an
ansatz is very similar to the calibration step performed in
phenomenological models. The work presented in this paper is only a very first, 
but necessary step into that direction.

\item Although this work is based on a single numerical simulation and a
particular star formation model, it already provides some interesting physical results. The
models fail to produce a significant color gradient in spiral
galaxies at low redshifts and are subject to a G-dwarf problem. Both problems
arise from the idealized boundary conditions, in
particular the lack of any late infall of gas. Other explanations include, but are
not limited to, the neglect of reddening due to dust.

\end{enumerate}

\acknowledgments

This work is partially supported by the  Deutsche 
Forschungsgemeinschaft under SFB 375-95 and 446 JAP-113/109/0. 
MS appreciates the hospitality at the Department of Astronomy, 
University of California at Berkeley, where part of this work has been 
performed.

\end{document}